\documentclass{article}
\usepackage{PRIMEarxiv}
\usepackage[utf8]{inputenc} 
\usepackage[T1]{fontenc}    
\usepackage{url}            
\usepackage{booktabs}       
\usepackage{amsfonts}       
\usepackage{nicefrac}       
\usepackage{microtype}      
\usepackage{lipsum}
\usepackage{fancyhdr}       
\usepackage{amsmath}
\usepackage{algpseudocode}
\usepackage{algorithm}
\usepackage{enumitem}
\usepackage[normalem]{ulem}
\usepackage{bm}
\usepackage{amssymb}
\usepackage{tabularx}
\usepackage{etoolbox}
\usepackage{color}
\usepackage{graphics}
\usepackage[dvipdfmx]{graphicx}
\graphicspath{{media}}     
\usepackage{array}
\usepackage{makeidx}
\usepackage{geometry}
\usepackage{booktabs}
\usepackage{multirow}
\usepackage{caption}
\usepackage{threeparttable}
\usepackage[table,xcdraw]{xcolor}
\def\Vec#1{\mbox{\boldmath $#1$}}  
\title{Extention of Bagging MARS with Group LASSO for Heterogeneous Treatment Effect Estimation

}

\author{
  Guanwenqing He \\
  Department of Medical Data-science\\
  Graduate School of Medicine\\
  Wakayama Medical University\\
  \texttt{mail} \\
   \And
  Ke Wan \\
  Department of Medical Data-science\\
  Graduate School of Medicine\\
  Wakayama Medical University\\
  \texttt{mail} \\
  \And
 Kazushi Maruo \\
  Department of Biostatistics\\
  Institute of Medicine\\
  University of Tsukuba \\
  \texttt{mail} \\
   \And
 Toshio Shimokawa \\
  Department of Medical Data-science\\
  Graduate School of Medicine\\
  Wakayama Medical University\\
  \texttt{mail} \\
}

\begin{document}
\maketitle

\begin{abstract}
Recent years, large scale clinical data like patient surveys and medical record data are playing an increasing role in medical data science. These large-scale clinical data, collectively referred to as "real-world data (RWD)". It is expected to be widely used in large-scale observational studies of specific diseases, personal medicine or precise medicine, finding the responder of drugs or treatments. Applying RWD for estimating heterogeneous treat ment effect (HTE) has already been a trending topic.  HTE has the potential to considerably impact the development of precision medicine by helping doctors make more informed precise treatment decisions and provide more personalized medical care. The statistical models used to estimate HTE is called treatment effect models. Powers {\it et al.} (2018)\cite{Powers2018} proposed a some treatment effect models for observational study, where they pointed out that the bagging causal MARS (BCM) performs outstanding compared to other models. While BCM has excellent performance, it still has room for improvement. In this paper, we proposed a new treatment effect model called shrinkage causal bagging MARS method to improve their shared basis conditional mean regression framework based on the following points: first, we estimated basis functions using transformed outcome, then applied the group LASSO method (Yuan and Lin, 2006)\cite{Yuan2006} to optimize the model and estimate parameters. Besides, we are focusing on pursing better interpretability of model to improve the ethical acceptance. We designed simulations to verify the performance of our proposed method and our proposed method superior in mean square error and bias in most simulation settings. Also we applied it to real data set ACTG 175 to verify its usability, where our results are supported by previous studies.
\end{abstract}


\section{Introduction}
With the development of electronic information technology and biomedical engineering, gathering and storing huge amounts of health-related data is faster and easier than ever before. Given this trend, large scale clinical data like patient surveys and medical record data are playing an increasing role in medical data science. These large-scale clinical data, collectively referred to as "real-world data (RWD)". It is expected to be widely used in large-scale observational studies of specific diseases, personal medicine or precise medicine, finding the responder of drugs or treatments(Schurman, 2019)\cite{schurman2019framework}. Although the randomized control trials (RCTs) are currently recognized as the highest quality evidence. However, RCTs are lengthy, expensive and often ethically challenging (Miksad{\it et al.}, 2019)(kravitz{\it et al.}, 2004)\cite{Miksad2019}\cite{kravitz2004evidence}. Additionally, randomized control trials are usually conducted in specific populations and settings that differ from the reality of the practical environment, which can cause problems of generalizability (Averitt{\it et al.}, 2020)(Greenfield{\it et al.}, 2007)\cite{Averitt2020}\cite{greenfield2007heterogeneity}. On the other hand, As observational data, real-world data can inform researchers in a efficient way saving time and cost while using real-world data yields answers that are relevant to a more general patient population (Sherman, 2016)(Kent{\it et al.}, 2018)\cite{Sherman2016}\cite{kent2018personalized}.

The application of RWD in precision medicine has already been a trending topic(Bica, 2021)\cite{bica2021real}. The purpose of reaching precision medicine is to help doctor select more personalized treatment options by offering quantitative evidence in order to improve the expected treatment effect. Researchers have developed some statistical estimating methods to estimate the nonrandom difference of individual in outcomes between two treatments based on covariates (eg.background, genetic information) (varadhan, 2013)\cite{varadhan2013framework}. Such difference is also called heterogeneous treatment effect (HTE). In some research it is called conditional average treatment effect (CATE).

Consider the HTE between an active treatment and an control treatment.  According to Neyman-Rubin’s counterfactual framework (Neyman, 1923; Rubin, 1974)\cite{neyman1923applications}\cite{rubin1974estimating}, The potential outcome of treatment group is $Y^{(1)}$, and the potential outcome of control group is $Y^{(0)}$, given covariate $\Vec{x}$, the HTE can be express as
\begin{equation}
{\tau}{(\bm{x})}=E[ Y^{(1)}-Y^{(0)} \mid \Vec{X} =\Vec{x}]. \nonumber
\end{equation}
\textcolor{red}{
But regarding observed outcome $Y$, we can only observe one of $Y^{(1)}$ or $Y^{(0)}$, we cannot have both treatment effects and control effects from a single individual at the same time. To figure this problem, treatment effect models are created to predict HTE . More recently, treatment effect models have been extended to the analysis of observational data, such as RWD. However, since observational data is often non-randomized, it can be difficult to establish causal relationships between the treatment effects and grouping. To address this issue, we need to use propensity adjustment to balance the confounding bias in treatment effect models(Rosenbaum and Rubin, 1983)(Pearl, 2009)\cite{Rosenbaum1983}\cite{pearl2009causality}. 
}

Researchers have proposed various treatment effect models\cite{Künzel2019}\cite{Athey2016}\cite{tian2014simple}\cite{athey2015machine}\cite{Wager2018}\cite{hill2011bayesian}. These models can be categorised into two main approaches: conditional mean regression methods and transformed outcome methods.  \cite{wendling2018comparing}\cite{Powers2018}\cite{HU2023102810}\cite{gutierrez2017causal}\cite{bica2021real}

In conditional mean regression, the estimation of heterogeneous treatment effect (HTE) involves two steps: first, fitting separate regression models for each group. The estimated regression model conditional on covariates for the treatment group $(t_i=1)$ is denoted as $\hat{f_1}(\Vec{x}_i)$, while the estimated regression model conditional on covariates for the control group $(t_i=0)$ is denoted as $\hat{f_0}(\Vec{x}_i)$. Secondly, the HTE is estimated by calculating the difference between the predicted outcomes from the regression models obtained in the first step. The HTE is estimated as
\begin{align}
\hat{\tau}{(\Vec{x}_i)}=\hat{f_1}(\Vec{x}_i)-\hat{f_0}(\Vec{x}_i)
\end{align}
 The estimation of HTE through conditional mean regression is heavily dependent on the accuracy of the regression models for the control and treatment groups. However, The true model is never known, and errors are always present. When we calculate the difference between the predicted outcomes of these two models, the errors in both models accumulate into the estimated HTE. It becomes challenging to figure out whether the estimated HTE is a result of the treatment effect or errors. Furthermore, conditional mean regression still requires propensity adjustment when confounding bias is present.

Transformed outcome regression is a method that convert the outcome $y_i$ and the treatment indicator $t_i$ into a single new outcome $z_i$ with the inverse probability treatment weight (IPTW)(Rosenbaum and Rubin, 1983)(Rosenbaum, 1987)\cite{Rosenbaum1983}\cite{rosenbaum1987model}. Transformed outcome can be represented by
\begin{align}
z_i=t_i\frac{y_i}{e(\Vec{x}_i)}+\left(1-t_i\right)\frac{{y}_i}{1-e\left(\Vec{x}_i\right)}, 
\end{align}
where $e\left(\Vec{x}_i\right)$ is the propensity score. An important characteristic of the transformed outcome is that, under the assumption of unconfoundedness assumption (Rosenbaum and Rubin, 1983)\cite{Rosenbaum1983}, the expectation of the transformed outcome conditional on covariate $\Vec{x}$ is the unbias estimation of HTE (Athey and Imbens, 2015)\cite{athey2015machine}. Therefore,the transformed outcome can then be used as the dependent variable in statistical models for estimating HTE. Transformed outcome avoid the problem of error accumulation. However, the introduction of propensity score leads to higher variance, especially when propensity scores are close to 0 and 1, which becomes particularly noticeable(Chesnaye, 2022)\cite{Chesnaye2022}. Transformed outcome regression provided a very intuitive approach to estimate HTE. However, it uses a single model for this estimation, unlike conditional mean regression which utilizes the difference between two models (\textbf{Eq.1}). This could pose an issue as the coefficients in the transformed outcome regression are typically smaller than those in  conditional mean regression. Consequently, the transformed outcome regression models might not effectively capture these differences.

To address the lack of transformed outcome methods and conditional mean regression methods, Powers {\it et al.} (2018) proposed shared basis conditional mean regression framework. Framework constructs a predictive models for heterogeneity in treatment effect at first, then estimate the outcomes of treatment group and control group separately using same basis functions (variables or their structures) constructed in the first step but different coefficient. Finally, the HTE is estimated by the difference between outcomes of treatment group and control group. As a result, the shared basis conditional mean regression could yield HTE with out the disturbance of varied selection of basis function, which can be challenging in metalearner framework methods. Therefore, the shared basis conditional mean regression framework suppose to reduce the bias of HTE.

\textcolor{red}{
Powers {\it et al.} (2018) proposed some shared basis conditional mean regression models to estimate HTE for both RCT data and observational data: pollinated transformed outcome (PTO) forests, causal multivariate adaptive regression splines (causal MARS). Powers {\it et al.} (2018) compared overall performance of multiple treatment effect models. It was ultimately concluded that the bagging causal MARS (BCM) surpassed that of the other models in high-dimensional data of both simulations and real data applications. BCM follows the shared basis conditional mean regression framework and for observational data they adopted propensity score stratification to reduce the confounding bias. Unlike Tree-based methods using the average treatment effect within terminal leaves to estimate HTE, BCM can provide precises estimation for each individual. This is even more crucial in the context of precision medicine research. However, the BCM method for observational data builds very complex models because of the application of propensity score stratification. Meanwhile, in BCM backward deletion does not affect the predictive ability of the model, but increases the computational cost, so the backward deletion part is excluded. Due to the above reasons, the model established by BCM has certain deficiencies in interpretability. 
}

We propose a novel approach: the shrinkage causal bagging MARS method to improve BCM in 2 main aspects:
\begin{enumerate}
    \item To adjust confounders, Powers {\it et al.} chose to use propensity score stratification in BCM. To avoid the disadvantage bring with stratification, we designed a strategy that build model to estimate transformed outcome to adjust confounders.
    \item In model construction, BCM make a adaptation of MARS to construct a shared basis conditional mean regression model (details in section 3). Our proposed method shrinkage causal bagging MARS follows the shared basis conditional mean regression framework in a totally different strategy. 

We are trying to rebalance the trade-off between model interpretability and predictive performance. By reducing the number of basis functions, we hope to make the model more interpretable while still maintaining its predictive power.
\end{enumerate}

We conduct extensive simulations to validate the performance of our proposed method. Then we compare it with several commonly used methods for HTE estimation. The results demonstrate superior performance in terms of accuracy and generalization ability, even in scenarios with complex data structures and varying treatment effects. We also validate our method on a real-world dataset, specifically the ACTG 175 dataset from the AIDS Clinical Trial, showcasing its applicability and reliability in practical settings.

The structure of this paper is as follows: Section 2 provide an overview of Multivariate adaptive regression splines (MARS) and the causal MARS model, respectively. In Section 3, the paper introduces a new method called shrinkage causal bagging MARS. The performance of various methods is evaluated through simulation studies in Section 4. Section 5 presents a real data application of proposed method.
At last we have a discussion of this study.

\section{Related Works}
Similar to BCM, our method proposed in this study is also built upon MARS, and it achieves superior predictive performance and interpretability compared to BCM. Therefore, in this section, we will briefly introduce MARS and BCM to help readers better understand our proposed method.

\subsection{Multivariate Adaptive Regression Splines (MARS) Method}
Multivariate adaptive regression splines (MARS) is a data analysis method proposed by Friedman (1991)\cite{Friedman1991}. It mainly deals with high-dimensional regression problems, and can be regarded as a generalization of stepwise linear regression, or an improvement of the performance of CART (classification and regression tree) (Breiman, 1984)\cite{Breiman2017} in regression problems. The MARS model can capture nonlinear relationships effectively, allowing for the modeling of complex interactions between features. And also as one of the generalized additive models, MARS has some interpretability and easy to be modified. 
Consider a dataset consists of $n$ individuals with $p$ covariates $\bm{x}_i=(x_{i1}, x_{i2},...,x_{ip})^\mathrm{T}$, where $i=1,2,...,n$. The model of the MARS method is given by
\begin{align}
f_{MARS}(\Vec{x}_i) &= \beta_0 + \sum_{m=1}^{M}{\beta_{m} {h_m(\Vec{x}_i)}}, \nonumber
\end{align}
where $M$ is the number of terms, $\beta_m$ is a regression parameter in the $m$-th term. $h_m(\Vec{x}_i)$ is $m$-th basis function:
\begin{align}
h_m(\Vec{x}_i) &= \prod_{k=1}^{K_m}{ \left[ s_{km}(x_{ij_{km}}-c_{km})\right]_+}, \nonumber
\end{align}
where$\left[ s_{km}(x_{ij_{km}}-c_{km})\right]_+$is the hinge function,
\begin{align}
\left[s_{km}\left(x_{ij_{km}}-c_{km}\right)\right]_+ &= \left\{\begin{matrix}
\max{\left(0,x_{ij_{km}}-c_{km}\right)}, & \ s_{km}=1 \\
  \max{\left(0,c_{km}{-x}_{ij_{km}}\right)}, & \ s_{km}=-1 \\
\end{matrix}\right.,
\end{align}
$K_m$ is the number of hinge functions of the $m$-th term, $s_{km}$ ($s_{km} \in \{+1,-1 \}$) is the sign function corresponding to the $k$-th hinge function of the $m$-th term, \textbf{Eq.3} called hinge function pair, $x_{ij_{km}}$ is the selected explanatory variables corresponding to the $k$-th hinge function of the $m$-th term of sample $i$. $c_{km}$ is selected cut-off values corresponding to the $k$-th hinge function of the $m$-th term of sample $i$.

MARS builds the model in two phases: the forward step and the backward deletion. For forward step, MARS starts with a model that contains only the intercept term of the mean response. A new interaction terms added to the MARS model is generated by multiplying an existing term with a new hinge function pair. Notice that, MARS add new terms to the model but not replace them. For each time, finds the hinge function pair that gives the maximum reduction in sum-of-squares residual error. The forward step usually builds an overfitted model. This process is repeated until the model reaches a predefined standard or a predefined number of basis functions.
Backward deletion removes inessential terms which contribute least to the model at each step. Backward stepwise uses generalized cross validation (GCV) to choose the best subset of model.


\subsection{Causal MARS (CM) and Bagging Causal MARS (BCM)}
The causal MARS model, proposed by Powers {\it et al.} (2018), is a treatment effect model based on multivariate adaptive regression splines (MARS). Consider $t_i$ denotes the treatment indicator. The model of causal MARS is given by
\begin{align}
f_{CM}(\Vec{x}_i,t_i) &= \sum_{m=1}^{M}{\beta_m^{(1)}h_m(\Vec{x}_i)}\mathbb{I}_{t_i=1}+\sum_{m=1}^{M}{\beta_m^{(0)}h_m(\Vec{x}_i)}\mathbb{I}_{t_i=0}\nonumber \\ 
&=\sum_{m=1}^{M}\beta_m^{(1)}\prod_{k=1}^{K_m} {\left[s_{km}\left(x_{ij_{km}}-c_{km}\right)\right]_+}\mathbb{I}_{t_i=1}+ \sum_{m=1}^{M}\beta_m^{(0)}\prod_{k=1}^{K_m} { \left[s_{km}\left(x_{ij_{km}}-c_{km}\right)\right]_+}\mathbb{I}_{t_i=0} \nonumber
\end{align}
where $\mathbb{I}_{t_i=1}$ and $\mathbb{I}_{t_i=0}$ is the group indicator function. if $t_i=1$,then $\mathbb{I}_{t_i=1}=1$ and $\mathbb{I}_{t_i=0}=0$, vice versa, $\beta_m^{(1)}$ is the coefficient of the treatment group for $m$-th term, and $\beta_m^{(0)}$ is the coefficient of the control group for $m$-th term. The causal MARS model constructs two MARS models by adding the same basis functions to fit the treatment and control groups. The criterion for selecting the best basis term involves evaluating the difference in the reduction of train error between a model with the same coefficient for the treatment and control groups and a model with different coefficients for the treatment and control groups. The basis term that results in the highest drop in train error when comparing the MARS model with different coefficients to the MARS model with the same coefficient is selected. See more details in Powers {\it et al.} (2018).

\textcolor{red}{
Power {\it et al.} (2018) suggested to use bagging procedure to reduce the variance of causal MARS, resulting in the bagging causal MARS (BCM). Bagging, short for bootstrap aggregating, is an ensemble learning technique that improves model performance and stability. It involves training multiple models on different subsets of the training data and combining their predictions. By combining the predictions of multiple models, bagging improves the overall performance and generalization ability of the ensemble model. Since the backward deletion and model selection didn't work very well in BCM, Power {\it et al.} omitted it in bagging causal MARS for saving computation. The optimal number of basic learners for the ensemble has not been assessed in BCM, which could potentially affect the model’s overall performance and accuracy (Chen and Jin, 2010)\cite{Chen2010}, while also making it overly complex.
}

For observational study data, they applied propensity score stratification to bagging causal MARS. The individuals $i$ are divided into $Q$ same stratus $\{s_q|q=1,...,Q\}$ by propensity score. Same basis function was selected in all stratum with a averaged criterion $\sum\Delta RSS_q$ but estimate the coefficient in each strata separately. The model formed as:
\begin{align}
f_{PBCM}(\Vec{x}_i,t_i)&=\sum_{q=1}^Q\mathbb{I}_{i\in s_q}\frac{1}{B}\sum_{b=1}^{B}(\sum_{m=1}^{M}{\beta_{mbq}^{(1)}h_{mb}(\Vec{x}_i)}\mathbb{I}_{t_i=1}+\sum_{m=1}^{M}{\beta_{mbq}^{(0)}h_{mb}(\Vec{x}_i)}\mathbb{I}_{t_i=0}),\nonumber
\end{align}
where $\beta_{mbq}^{(1)}$ is treatment group coefficient of $m$-th term in causal MARS model corresponding to stratus $s_q$ in bootstrap sample $b$ and $\beta_{mbq}^{(0)}$ is control group coefficient of $m$-th term in causal MARS model corresponding to stratus $s_q$ in bootstrap sample $b$. $h_{mb}(\Vec{x}_i)$ is the $m$-th term in causal MARS model corresponding to bootstrap sample $b$, it remains consistent across all strata.

However, as the number of stratum increases, the population is reduced within each stratum. This reduction in population size can result in increased model instability, particularly when the total population is not sufficiently large (Linden, 2017)\cite{Linden2017}. Additionally, if the assignment of patients in the treatment group and the control group is imbalanced within specific strata, it further exacerbates the issue. Despite this, confoundness within strata are ignored (Streiner and Norman, 2012)\cite{Streiner2012}. Additionally, the combination of multiple models for each stratum also increases the complexity of the model and it makes the model harder to interpret. 

\section{Shrinkage Causal Bagging MARS (SCBM)}\label{sec3}

\textcolor{red}{
In certain medical scenarios, machine learning has a fundamental trade off. The predictive outcome alone may offer only a partial solution, but researchers in the context of medical research usually has more curious on the reasons behind predicted outcomes. However, the commonly used treatment effect models models (include BCM) currently are mostly black box models, focusing on predictive performance rather than interpretability. To address this, the interpretable machine learning gains momentum. Interpretability allows us to extract this additional knowledge captured by the model (Doshi-Velez and Kim, 2017)\cite{doshi2017towards}. Moreover, in real-world applications, where safety, ethic and rigorous testing are required, interpretability becomes more essential (molnar, 2020)\cite{molnar2020interpretable}. 
}

\textcolor{red}{
In this study, we are trying to balance the trade-off between model interpretability and predictive performance. our target is to generate an intuitive and simple interpretable model which has a generalized additive model structure and can be interpreted to some extent by analyzing the coefficients of the corresponding base functions and variables. The model proposed in this study aims to improve interpretability while still maintaining its predictive power. In section 2.2, we introduced the BCM model and mentioned the advantages and several issues with it. One issue arises from propensity score stratification, while another pertains to the BCM model's lack of backward deletion. Those issues lead to a overly compelx model and potentially make the model instability. 
}

\textcolor{red}{
Regarding the problem of propensity score stratification, there are two common alternative solutions: propensity score matching and the IPTW method. In propensity score matching, unmatched individuals are often excluded, making it less preferable in most cases. On the other hand, IPTW methods can effectively summarize multiple confounders into a single variable, retaining most individuals and enhancing data utilization. Regrettably, within the context of BCM, IPTW is not applicable. While BCM incorporates the treatment indicator as a crucial element in building the model, IPTW converts this indicator into a new outcome variable, making it incompatible with BCM. In this study, we use transformed outcome to adjust confounders. The transformed outcome method adjusts confounders by giving a unbiased estimation of HTE with propensity score. Therefore, generally,  transformed outcome method can also be considered an IPTW method. 
}

\textcolor{red}{
Regarding the problem of backward deletion, researchers \cite{BOLONCANEDO20191} \cite{polikar2012ensemble} has found out that using ensemble leaning to do feature selection can evidently improve the predict accuracy. We use group LASSO (Least Absolute Shrinkage and Selection Operator) (Yuan and Lin, 2006)\cite{Yuan2006} regression to estimate the optimized model by removing the inessential terms, and regularization, while ensuring consistency of the basis functions between the two groups, as mentioned in shared basis conditional mean regression.  While Chen and Jin (2010)\cite{Chen2010} pointed out that bagging with shrinkage estimation (e.g. LASSO) has a higher prediction accuracy than ordinary bagging. So we also applied ordinary LASSO to feature selection, and the Simulation reasult is shown in \textbf{Section 4} (In this section, we do not discuss this method).
}

\subsection{Methodology}
\textcolor{red}{
The model construction consists of three steps: basis function generation, coefficient estimation, HTE estimation. Figure 1 provides a brief overview of the model algorithm. 
}
\begin{figure}[!t]
\begin{center}
\includegraphics[width=1\linewidth]{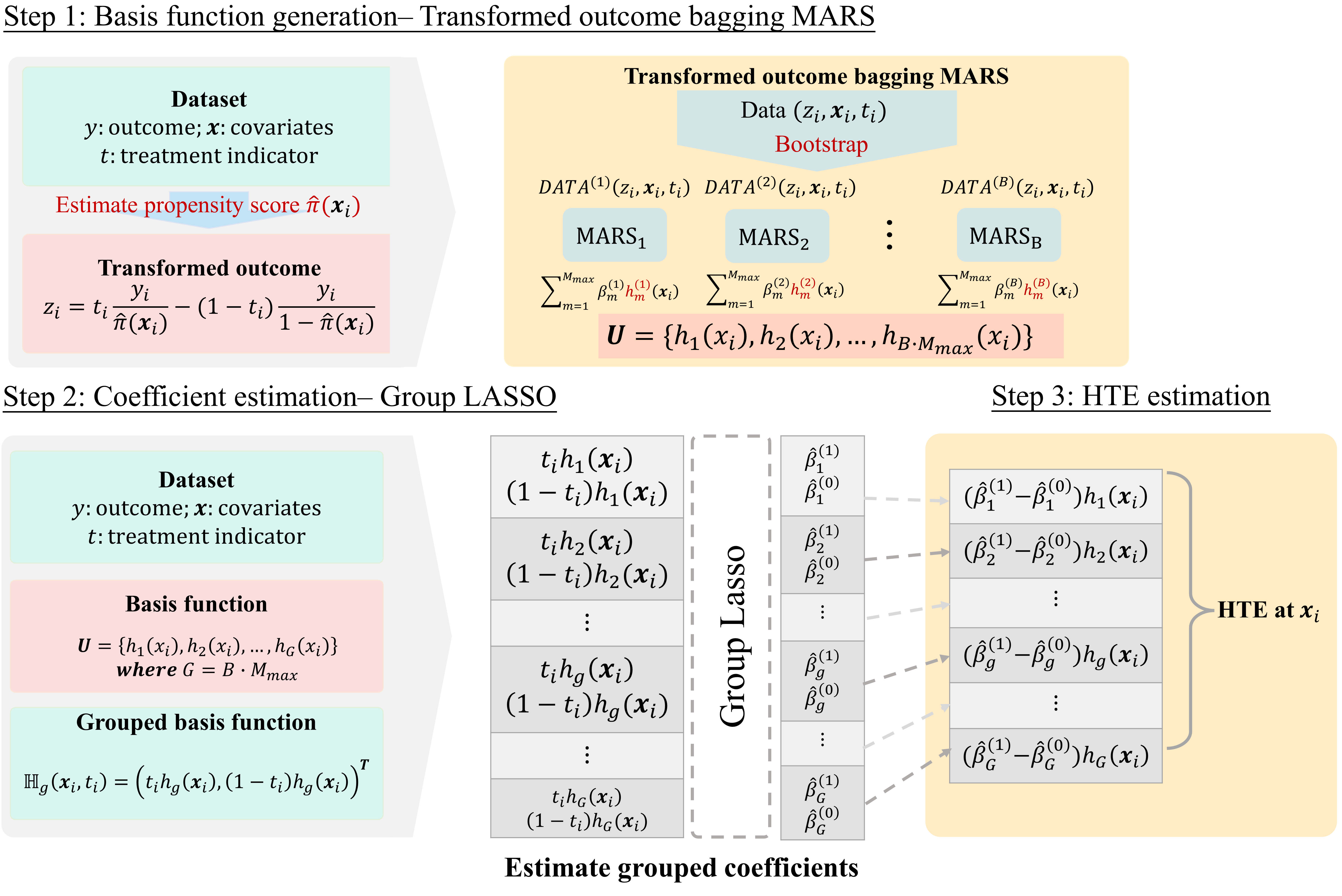}
\caption{Summarized algorithm of the proposed method.}
\label{fig:scbm}
\end{center}
\end{figure}
\begin{algorithm}[t]
\caption{Shrinkage Causal Bagging MARS Algorithm}\label{alg2}
\begin{algorithmic}[1]
\Statex \textbf{Input: }$ Data\{(\bm{x}_i,t_i,y_i)\}_{i=1}^N, M_{max}, K_{max}, B $
\Statex $\texttt {// Basis function collection $\bm{U}$ is generated as follows}$
\Statex \textbf{Initialize: }$\bm{U}\leftarrow\phi, z_i=t_i\displaystyle\frac{y_i}{e(\Vec{x}_i)}+\left(1-t_i\right)\displaystyle\frac{{-y_i}}{1-e\left(\Vec{x}_i\right)}.$
\For {$\text{bootstrap sample}\ Data^{(b)}\{(\bm{x}_i,t_i,z_i)\}^{(b)}, \text{where}\ b\in 1,\cdots, B$}
\State $h^{(b)}_1(\bm{x}_i) \leftarrow 1; M \leftarrow 2. $ 
\State $ \textbf{Loop until } M \geq M_{max} $
\State $ LOF^* \leftarrow \infty $
\For{$m=1$ to $M-1$}
\For{$j\notin\{j(k,m)|1\leq k \leq K_{max}\}$}
\For{$c\in\{x_{ij}|h_m^{(b)}(\bm{x}_i)>0\}$}
\State $LOF\leftarrow\min_{\beta_1,...,\beta_{M+1}}\sum_{i=1}^n\big(z_i-\sum_{v=1}^{M-1}\beta_vh^{(b)}_v(\bm{x}_i)+$
\Statex \hspace{6cm}    $\beta_Mh^{(b)}_m(\bm{x}_i)[+(x_{ij}-c)]_++\beta_{M+1}h^{(b)}_m(\bm{x}_i)[-(x_{ij}-c)]_+\big)$
\State $\textbf{if }LOF<LOF^*,\text{then }LOF^*\leftarrow LOF;m^*=m;j^*=j;c^*=c \textbf{ end if}$
\EndFor
\EndFor
\EndFor
\State $h^{(b)}_M(\bm{x}_i)\leftarrow h^{(b)}_{m^*}(\bm{x}_i)[+(x_{ij^*}-{c^*})]_+$
\State $h^{(b)}_{M+1}(\bm{x}_i)\leftarrow h^{(b)}_{m^*}(\bm{x}_i)[-(x_{ij^*}-{c^*})]_+$
\State $M\leftarrow M+2$
\State $\bm{U}=\bm{U}\cup \{h^{(b)}_M(\bm{x}_i),h^{(b)}_{M+1}(\bm{x}_i)\}$
\State $\textbf{end loop}$
\EndFor
\Statex \texttt {// Reorder the index of basis function collection $\bm{U}$}
\State $\bm{U}=\{h_1(\bm{x}_i),h_2(\bm{x}_i),...,h_{B\cdot M_{max}}(\bm{x}_i)\}$
\State $\mathbb{H}_g(\Vec{x}_i,t_i )=((h_g(\bm{x}_i)\mathbb{I}_{t_i=1}),(h_g(\bm{x}_i)\mathbb{I}_{t_i=0}))^\mathrm{T}$
\Statex $\texttt{// The group LASSO objective function is given by}$
\State $LOF_{gl}=\sum_{i=1}^n \big(y_i-\sum_{g=1}^{B\cdot M_{max}}\Vec{\beta}_g^\mathrm{T}\mathbb{H}_g(\Vec{x}_i,t) \big)^2 + \lambda_{gl}\sum_{g=1}^{B\cdot M_{max}}\|\Vec{\beta}_{g}\|_2,$ $\text{where $\Vec{\beta}_g=(\beta_g^{(1)},\beta_g^{(0)})^\mathrm{T}$}$
\State $\{\hat{\Vec{\beta}_g}|g=1,...,M_{max}\}=\arg\min_{\{\Vec{\beta}_g|g=1,...,M_{max}\}}LOF_{gl}$
\Statex $\texttt {// Final model of SCBM is given by}$
\State $\widehat{\tau}\left(\Vec{x}_i\right)=\sum_{g=1}^{B\cdot M_{max}}{{\hat{\Vec{\beta}}}_g^\mathrm{T}\mathbb{H}_g(\Vec{x}_i,1)}-\sum_{g=1}^{B\cdot M_{max}}{{\hat{\Vec{\beta}}}_g^\mathrm{T}\mathbb{H}_g(\Vec{x}_i,0)}=\sum_{g=1}^{B\cdot M_{max}}(\beta_g^{(1)}-\beta_g^{(0)})h_g(\bm{x}_i).$
\Statex \textbf{Output: }$ \widehat{\tau}\left(\Vec{x}_i\right) $
\end{algorithmic}
\end{algorithm}

\paragraph{Step 1 --- Basis function generation}
In this step, we are going to generate basis functions to obtain HTE. The constructed basis functions are supposed to indicate the HTE. We chose MARS as the basis function generator. To adjust confounders while obtain the HTE, we applied transformed outcome method, which combine the treatment indicator and original outcome into a new outcome with propensity score. And the use of bootstrap sampling to reduces variance and improves model stability.

The algorithm of the basis function generation is presented in $\mathbf{Algorithm}\ \ref{alg2}$ line 1 to line 19.
Given a dataset $Data\{(\bm{x}_i,t_i,y_i)\}$. Transformed outcome $z_i$ is estimated by \textbf{Eq.2}. Given Hyper parameters maximum number of term in MARS $M_{max}$ and maximum number of interaction degree in MARS $K_{max}$, and number of bootstrap samples $B$. Basis function collection is initialized as $\bm{U}\leftarrow\phi$, where $\phi$ is null set.

The algorithm applies bootstrap sampling to create $B$ bootstrap samples $Data^{(b)}\{(\bm{x}_i,t_i,z_i)\}_{i=1}^{(b)}$ in line 1, where $b=1,...,B$. Each bootstrap sample consists of covariates $\bm{x}_i$, treatment indicator $t_i$, and transformed outcome $z_i$. In line 2, the intercept term of MARS $h^{(b)}_1(\bm{x}_i)$ is set to $1$, terms counter $M$ is set to $2$. In our proposed method, the propensity score is estimated with random forest for a better accuracy and less model dependency (zhao, 2016)\cite{Zhao2016}.

\textcolor{red}{
Let transformed outcome $z_i$ as the outcome and matrix of covariate $\bm{x}_i$ as input, we construct a transformed outcome bagging-MARS model as in line 2 to line 18.
 \begin{enumerate}[label=\textcircled{\Alph*}]
     \item Until $M \geq M_{max}$ iterate $m \leq M-1$, traverse every sample of $Data^{(b)}$ find a certain variable of a sample as cut points to construct the hinge function pair like \textbf{Eq.3}. Then the new terms is construct by multiplying hinge function pair to any of existing term in the MARS model (include the intercept one). Notice that, in the new terms, the cut points variable cannot appear twice(line 6). Meanwhile the cut point sample should not make the existing term values about to be multiplied equal to 0. (line 7).     
     \item The optimal terms is selected by minimizing the LOF (loss of function) in line 8. The regression coefficients are estimated by minimizing the LOF.  After selecting the optimal terms, the new terms are added to the basis function collection $\bm{U}$.  The terms counter $M$ is increased by 2.
 \end{enumerate}
 \textcircled{A} and \textcircled{B} are repeated until the specified number of iterations $M_{max}$ is reached. The transformed outcome bagging-MARS model of bootstrap sample $b$ is given by
\begin{align}
f_{TBM}^{(b)}\left(\Vec{x}_i\right)=\sum_{m=1}^{M_{max}} \hat{\beta}^{(b)}_m h^{(b)}_m(\bm{x}_i) \nonumber
\end{align}
where $\hat{\beta}^{(b)}_m$ is the coefficient of $m$-th term of bootstrap sample $b$, $h^{(b)}_m(\bm{x}_i)$ represents the $m$-th term of bootstrap sample $b$. $h^{(b)}_m(\bm{x}_i)=\prod_{k=1}^{K_{m}^{(b)}}\left[s_{kmb}\left(x_{i{j_{kmb}}}-c_{kmb}\right)\right]_+$, where $K_{m}^{(b)}$ is the degree of interaction of $m$-th term of bootstrap sample $b$.
After contract MARS model for every bootstrap sample, all the basis functions are combined to form a new collection $\bm{U}$: 
\begin{align}
\bm{U}=\{h_1(\bm{x}_i),h_2(\bm{x}_i),...,h_{B\cdot M_{max}}(\bm{x}_i)\}
\end{align}
}

\textcolor{red}{
\paragraph{Step 2 --- Coefficient estimation}
We have generated a treatment effect model in \textbf{Step 1} already, but as we discussed in introduction, transformed outcome bagging-MARS has high variance and the variance is also highly associated with number of samples and main effect, which has been proved in Powers {\it et al.} (2018). So we prefer to use conditional mean regression in \textbf{Step 2} and using the basis functions $\bm{U}$ gathered in \textbf{Step 1}. Refer to \textbf{Eq. 1}, we can estimate the HTE by the difference of two models. And, we use the group LASSO method in regression to estimate the coefficient of basis functions for diffenent treatment group. As a shrinkage methods, LASSO help to address the problem of overfitting and improved model generalization. By punish extreme parameter values or overly complex models, shrinkage methods promote a balance between model complexity and data fitting while improving the model interpretability. Group LASSO is an extension of LASSO to handle the variables with group structure. In our proposed method, group LASSO do the basis function selection and maintain the basis function same for treatment group and control group which can avoid the bias cause by the defences between groups in basis function selection when estimating HTE. By doing this, we can yield a model to estimate HTE while avoiding the disadvantage of transformed outcome model and conditional mean regression.
}

The algorithm of the coefficient estimation is presented in $\mathbf{Algorithm}\ \ref{alg2}$ line 20 to line 22.
After constructing the basis functions collection in \textbf{Eq.3}, we construct the $g$-th grouped basis function vector $\mathbb{H}_g(\Vec{x}_i,t_i )=((h_g(\bm{x}_i)\mathbb{I}_{t_i=1}),(h_g(\bm{x}_i)\mathbb{I}_{t_i=0}))^\mathrm{T}$ which has two same basis functions $h_g(\bm{x}_i)$ multiplied with group indicators $\mathbb{I}_{t_i=1}$ and $\mathbb{I}_{t_i=0}$. The objective function $LOF$ for group LASSO is defined, which consists of the squared prediction errors and the group LASSO penalty. 
The objective function $LOF$ for SCBM is defined as 
\begin{align}
LOF_{gl}=\sum_{i=1}^n \big(y_i-\sum_{g=1}^{B\cdot M_{max}}\Vec{\beta}_g^\mathrm{T}\mathbb{H}_g(\Vec{x}_i,t) \big)^2 + \lambda_{gl}\sum_{g=1}^{B\cdot M_{max}}\|\Vec{\beta}_{g}\|_2,
\end{align}
where $\Vec{\beta}_g=(\beta_g^{(1)},\beta_g^{(0)})^\mathrm{T}$,  $\| \cdot \|_2 $ is the $l^2$-norm ,and $\lambda_{gl}$ is shrinkage degree parameter. The regression coefficients collection $\hat{\Vec{\beta}}=\{\hat{\Vec{\beta}}_g|g=1,...,M_{max}\}$ is estimated by minimizing the objective function $LOF$ in \textbf{Eq.4}, resulting in the optimal regression coefficient estimates. 

\paragraph{Step 3 --- HTE estimation}
The HTE is estimated by comparing the outcomes of the group LASSO model between the treatment and control groups.The algorithm of the coefficient estimation is presented in $\mathbf{Algorithm}\ \ref{alg2}$ line 23.
The final SCBM model predicts the causal effects by calculating the difference between the predicted results for the treatment group and the control group. The causal effect is given by 
\begin{align}
\widehat{\tau}\left(\Vec{x}_i\right)&=\sum_{g=1}^{B\cdot M_{max}}{{\hat{\Vec{\beta}}}_g^\mathrm{T}\mathbb{H}_g(\Vec{x}_i,1)}-\sum_{g=1}^{B\cdot M_{max}}{{\hat{\Vec{\beta}}}_g^\mathrm{T}\mathbb{H}_g(\Vec{x}_i,0)}=\sum_{g=1}^{B\cdot M_{max}}(\beta_g^{(1)}-\beta_g^{(0)})h_g(\bm{x}_i)\nonumber\\
&=\sum_{g=1}^{B\cdot M_{max}}(\beta_g^{(1)}-\beta_g^{(0)})\prod_{k=1}^{K_{g}}\left[s_{kg}\left(x_{i{j_{kg}}}-c_{kg}\right)\right]_+,\nonumber
\end{align}

\section{Data Simulation}\label{sec4}
In this section, we evaluated performance of several methods for heterogeneous treatment effect estimation in both randomized controlled trial and observational studies. We compare our proposed methods to causal forest, PTO forest, causal MARS, virtual twins. Our goal of this simulation is to make a contrast on the ability of our proposed method shrinking causal MARS with other method in estimating heterogeneous treatment effect in many scenarios.
\subsection{Simulation Setting}
The population of patients is $n$, for each patient, there are $p$ variables observed. The simulation settings are as following:
\begin{align}
\mathrm{Predictor\ variables}\ 
x_{ij}\sim\left\{\begin{matrix}
N\left(0,1\right), & \  j\mathrm{\ is\ odd\ number}\  \\
Ber\left(\frac{1}{2}\right),&  j\mathrm{\ is\ even\ number}\ \\ 
\end{matrix}\right.\nonumber,
\end{align}

\begin{align}
\mathrm{Treatment\ variable}\ 
t_i\sim Ber\left(e\left(\Vec{x}_i\right)\right)\nonumber,
\end{align}
Where $e\left(\Vec{x}_i\right)$ is the propensity score function which defined as
\begin{align}
e\left(\Vec{x}_i\right)\sim\left\{\begin{matrix}
\frac{1}{2}, & \ \mathrm{RCT\ situation}  \\
\frac{\exp\left\{\mu\left(\Vec{x}_i\right)-\tau(\Vec{x}_i)/2\right\}}{1+\exp\left\{\mu\left(\Vec{x}_i\right)-\tau(\Vec{x}_i)/2\right\}},&  \mathrm{Observational\ study\ situation} \\ 
\end{matrix}\right.\nonumber,
\end{align}

\begin{align}
\mathrm{Outcome}\ 
y_i\sim N\left(\mu\left(\Vec{x}_i\right)+\left(t_i-\frac{1}{2}\right)\tau\left(\Vec{x}_i\right),1\right),\nonumber
\end{align}
where the main effect function is $\mu\left(\Vec{x}_i\right)$ and treatment effect function is $ \tau\left(\Vec{x}_i\right) $. To evaluate the predictive accuracy of the models, the performance metrics we adopt are mean squared error (MSE) and absolute bias.
\begin{align}
\text{MSE} = \frac{1}{n}\sum_{i=1}^{n}(y_i-\hat{f}(\Vec{x_i}))^2\nonumber\\
\text{absolute bias} = |\frac{1}{n}\sum_{i=1}^{n}(\hat{f}(\Vec{x_i}) - y_i)|\nonumber
\end{align}

There are 12 simulations. Simulation 1 to 6 is set to simulate the RCT study, and simulation 7 to 12 is set to simulate the observational study. All the simulations were run 250 times. In each simulation, the main effect function $\mu\left(\Vec{x}_i\right)$, treatment effect function $\tau\left(\Vec{x}_i\right)$ are defined as $\mathbf{Table}$ \ref{tb1}  and there are 3 sets of sample size $n$ and number of covariates $p$: setting.1: $n$ =200, $p$ =100, setting.2: $n$ =500, $p$ =200, setting.3: $n$ =500, $p$ =400.

\begin{table}[ht]
\centering
\caption{ Model of main effect and treatment effect.}
\label{tb1}
    \begin{threeparttable}
        \begin{tabular}{cccc}
            \hline
        
            \multicolumn{2}{c}{Simulation No.}                  & \multirow{2}{*}{Model}   &  \multirow{2}{*}{Functional form} \\
                     RCT         &       OBS            &           &      \\
                              \midrule
            \multirow{2}{*}{1} & \multirow{2}{*}{7} &        $\mu\left(\Vec{x}_i\right)=2x_{i1}-4$       &  linear  \\
                              &                   &          $  \tau\left(\Vec{x}_i\right)=0  $   &  constant \\
                              \midrule
            \multirow{2}{*}{2} & \multirow{2}{*}{8} &           $\mu\left(\Vec{x}_i\right)=5\mathbb{I}_{\left\{x_{i1}>1\right\}}-5$     & piecewise  \\
                              &                   &        $\tau\left(\Vec{x}_i\right)=4\mathbb{I}_{\left\{x_{i1}>1\right\}}\mathbb{I}_{\left\{x_{i3}>1\right\}}+4\mathbb{I}_{\left\{x_{i5}>1\right\}}\mathbb{I}_{\left\{x_{i7}>1\right\}}+2x_{i8}x_{i9}$        &  piecewise interactive \\
                              \midrule
            \multirow{2}{*}{3} & \multirow{2}{*}{9} &          $\mu\left(\Vec{x}_i\right)=\frac{1}{2}\left(x_{i1}^2+x_{i2}+x_{i3}^2+x_{i4}+x_{i5}^2+x_{i6}+x_{i7}^2+x_{i8}+x_{i9}^2-11\right)$       &  additive \\
                              &                   &           $\tau\left(\Vec{x}_i\right)=\frac{1}{\sqrt2}\left(f_1\left(x\right)^{*}+f_2\left(x\right)^{**} \right)$      &  interactive\\
                              \midrule
            \multirow{2}{*}{4} & \multirow{2}{*}{10} &         $\mu\left(\Vec{x}_i\right)=4\mathbb{I}_{\left\{x_{i1}>1\right\}}\mathbb{I}_{\left\{x_{i3}>1\right\}}+4\mathbb{I}_{\left\{x_{i5}>1\right\}}\mathbb{I}_{\left\{x_{i7}>1\right\}}+2x_{i8}x_{i9}$     & piecewise interactive \\
                              &                   &           $\tau\left(\Vec{x}_i\right)=f_2\left(\Vec{x}_i\right)^{**} $   &  linear\\
                              \midrule
            \multirow{2}{*}{5} & \multirow{2}{*}{11} &         $\mu\left(\Vec{x}_i\right)=f_1\left(\Vec{x}_i\right)^{*}$     & piecewise interactive  \\
                              &                   &           $\tau\left(\Vec{x}_i\right)=\sin{\left(\pi x_{i1}x_{i3}\right)}+2\left(x_{i5}-0.5\right)^2+x_{i7}+0.5x_{i9}$   &  interactive\\   
                              \midrule
            \multirow{2}{*}{6} & \multirow{2}{*}{12} &         $\mu\left(\Vec{x}_i\right)=0.5-0.1/(1+e^{-x_{i1}})+0.1\sin(x_{i3})-0.1x_{i5}^2-0.2x_{i5}-0.1x_{i7}^2$     & additive\\
                              &                   &           $\tau\left(\Vec{x}_i\right)=-0.2+0.5\sin(\pi x_{i1}x_{i3})+0.2/(1+e^{-x_{i5}})+0.2x_{i2}+0.3x_{i4}$   &  interactive \\
        \hline
        \end{tabular}
        \begin{tablenotes}
        \footnotesize
            \item[* ]  $f_1\left(\Vec{x}_i\right)=x_{i2}x_{i4}x_{i6}+{2x}_2x_{i4}(1-x_{i6})+{3x}_2{(1-x}_4)x_{i6}+4x_{i2}{(1-x}_4)(1-x_{i6})+ 5(1-x_{i2})x_{i4}x_{i6} +6(1-x_{i2})x_{i4}(1-x_{i6})+7(1-x_{i2}){(1-x}_4)x_{i6}+8(1-x_{i2}){(1-x}_4)(1-x_{i6})$  
            \item[**] $f_2\left(\Vec{x}_i\right)=x_{i1}+x_{i3}+x_{i5}+x_{i7}+x_{i8}+x_{i9}-2$ 
        \end{tablenotes}
    \end{threeparttable}
\end{table}

\begin{figure}[!ht]
\begin{center}
\includegraphics[width=1.0\linewidth]{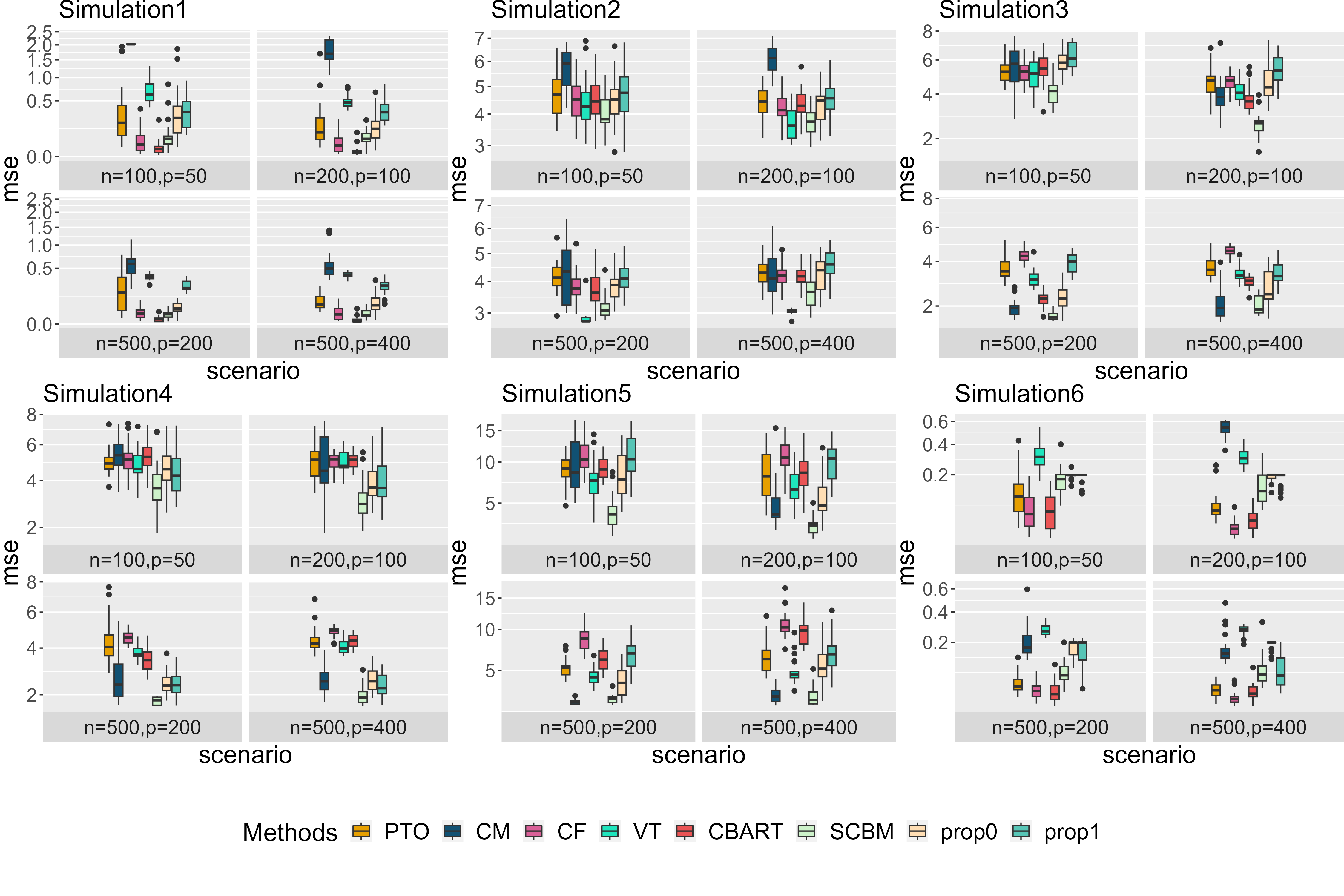}
\caption{MSE of simulations 1 to 6, The x-axis is the number of covariates and subjects in the
simulation dataset for the corresponding scenario. The y-axis is the mean square error (mse) value of 8
different methods, including PTO = pollinated transformed outcome forest, BCM = bagged causal MARS, CF = causal forest, VT = virtual twins,CM = causal MARS, BART = Bayesian additive regression tree, SCBM = Shrinkage Causal Bagging MARS (proposed method), prop0 = transformed outcome bagging-
MARS(with LASSO in each ensemble samples), prop1 = transformed outcome bagging-
MARS(with LASSO in each ensemble samples and ridge for all basis functions).}
\label{fig:11}
\end{center}
\end{figure}
\begin{figure}[!ht]
\begin{center}
\includegraphics[width=1.0\linewidth]{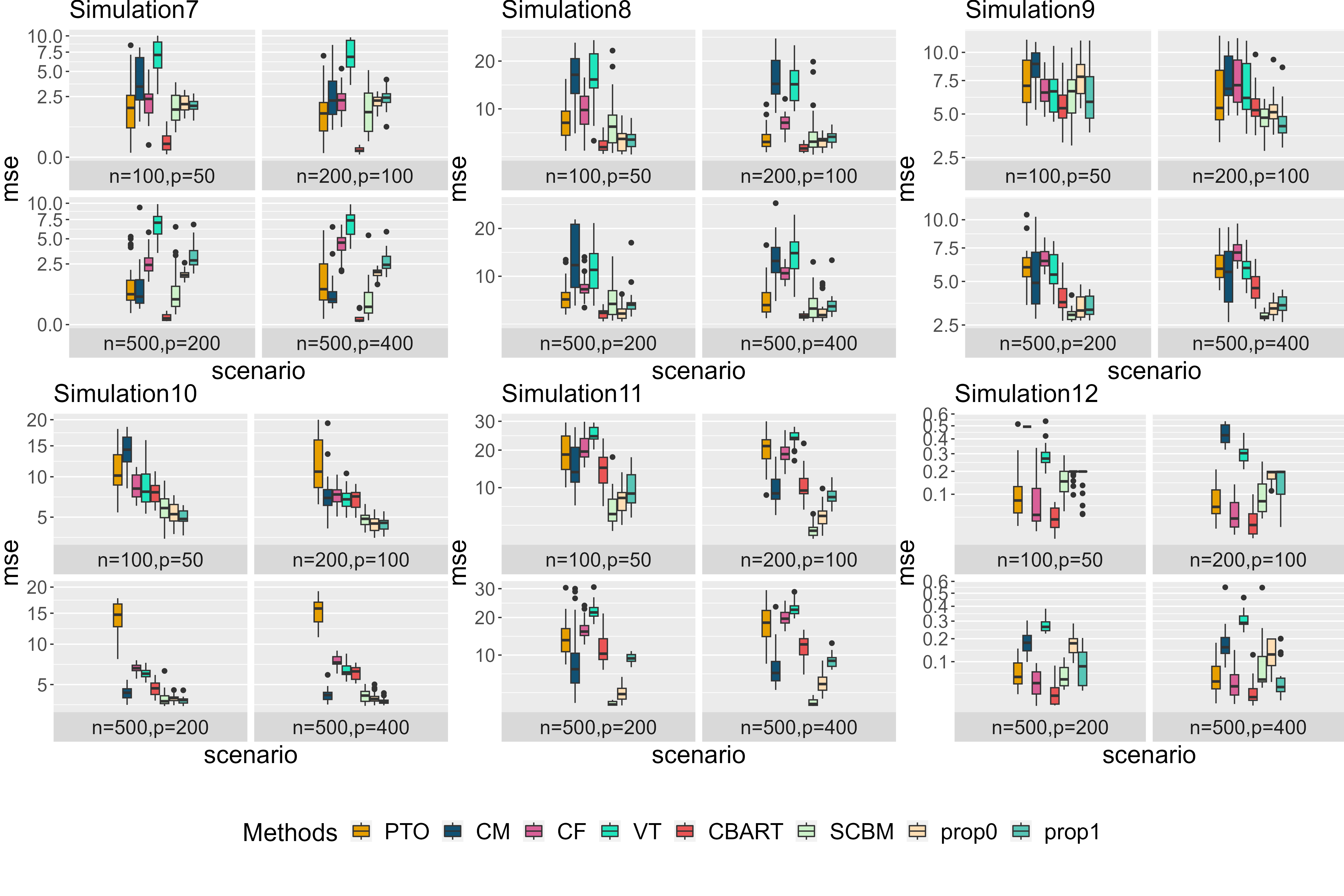}
\caption{MSE for simulations 7 to 12, The x-axis is the number of covariates and subjects in the
simulation dataset for the corresponding scenario. The y-axis is the mean square error (mse) value of 8
different methods, including PTO = pollinated transformed outcome forest, BCM = bagged causal MARS, CF = causal forest, VT = virtual twins,CM = causal MARS, BART = Bayesian additive regression tree, SCBM = Shrinkage Causal Bagging MARS (proposed method), prop0 = transformed outcome bagging-
MARS(with LASSO in each ensemble samples), prop1 = transformed outcome bagging-
MARS(with LASSO in each ensemble samples and ridge for all basis functions).}
\label{fig:12}
\end{center}
\end{figure}
\begin{figure}[!ht]
\begin{center}
\includegraphics[width=1.0\linewidth]{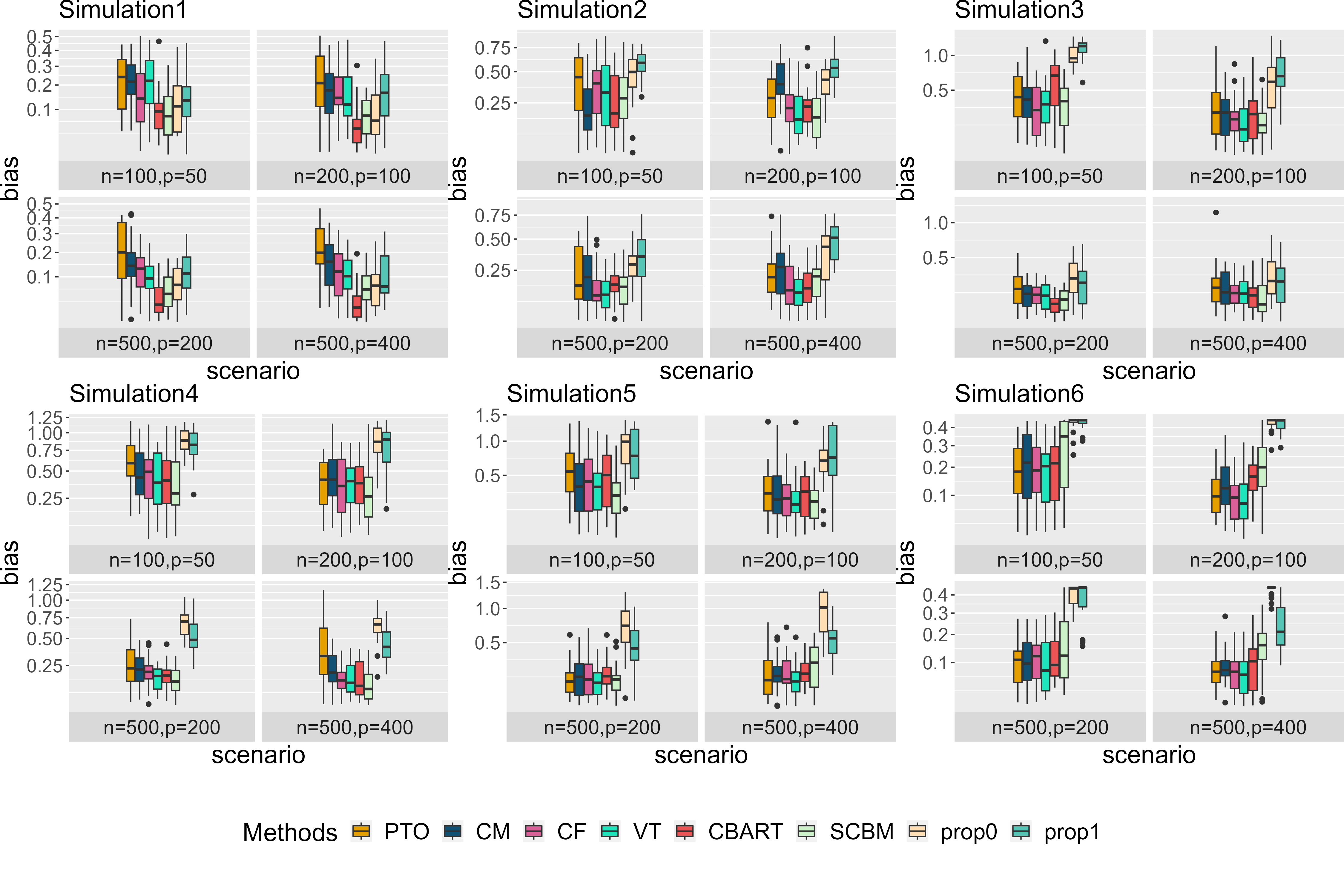}
\caption{Bias of simulations 1 to 6, The x-axis is the number of covariates and subjects in the
simulation dataset for the corresponding scenario. The y-axis is the absolute bias (bias) value of 8
different methods, including PTO = pollinated transformed outcome forest, BCM = bagged causal MARS, CF = causal forest, VT = virtual twins,CM = causal MARS, BART = Bayesian additive regression tree, SCBM = Shrinkage Causal Bagging MARS (proposed method), prop0 = transformed outcome bagging-
MARS(with LASSO in each ensemble samples), prop1 = transformed outcome bagging-
MARS(with LASSO in each ensemble samples and ridge for all basis functions).}
\label{fig:21}
\end{center}
\end{figure}
\begin{figure}[!ht]
\begin{center}
\includegraphics[width=1.0\linewidth]{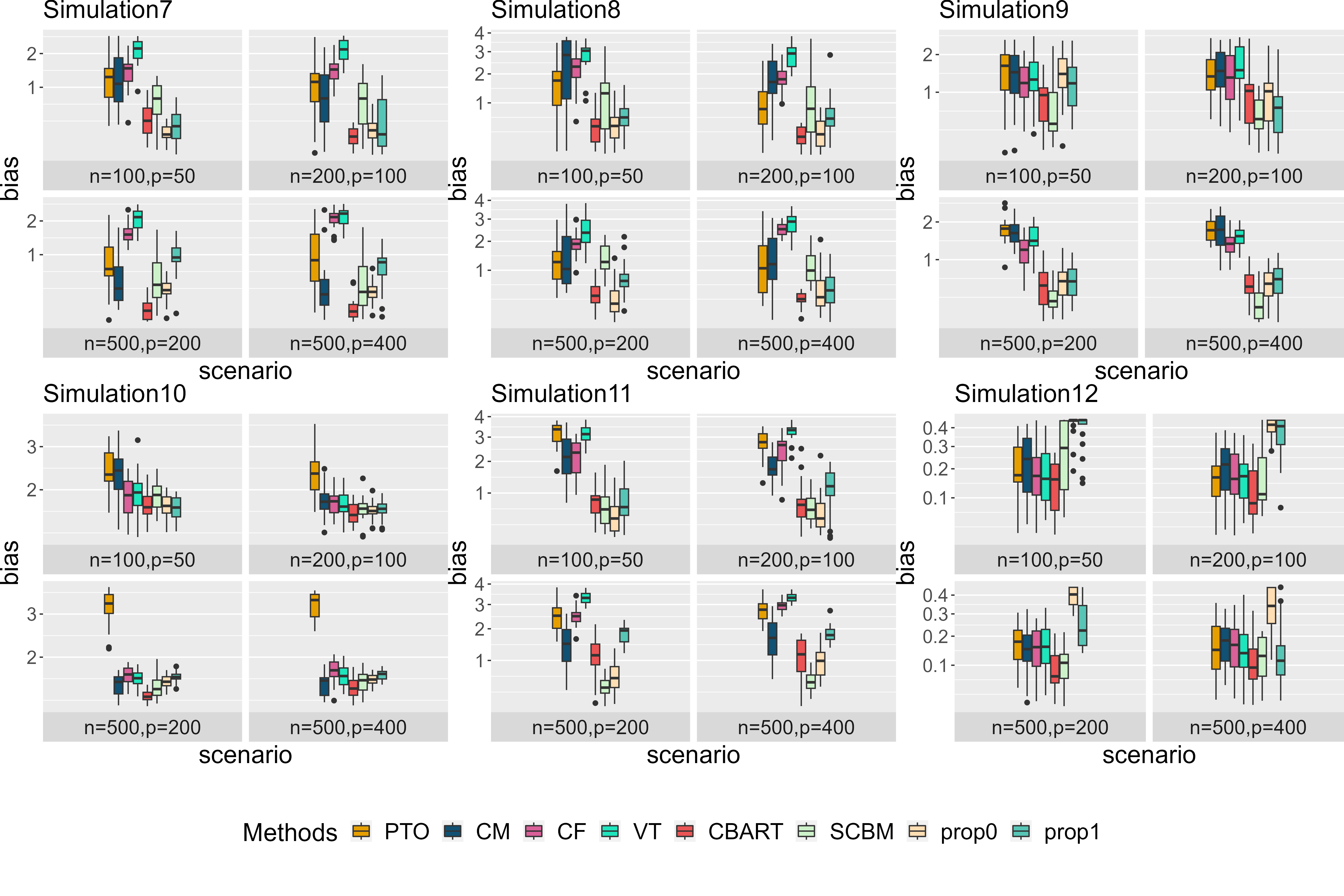}
\caption{Bias for simulations 7 to 12, The x-axis is the number of covariates and subjects in the
simulation dataset for the corresponding scenario. The y-axis is the absolute bias (bias) value of 8
different methods, including PTO = pollinated transformed outcome forest, BCM = bagged causal MARS, CF = causal forest, VT = virtual twins,CM = causal MARS, BART = Bayesian additive regression tree, SCBM = Shrinkage Causal Bagging MARS (proposed method), prop0 = transformed outcome bagging-
MARS(with LASSO in each ensemble samples), prop1 = transformed outcome bagging-
MARS(with LASSO in each ensemble samples and ridge for all basis functions).}
\label{fig:22}
\end{center}
\end{figure}

\subsection{Simulation Result}  
Figure 2 is the result of MSE of RCT study simulations 1 to 6. In simulation 1, which has a constant treatment effect, the causal BART dominated in all scenarios, and our proposed method performed second to third place. In simulation 2, which has a piecewise interactive treatment effect, VT performed best in all scenarios and our proposed method also looks good. In simulation 3, which has an additive treatment effect our proposed method dominate in scenario of $n=500$, and causal MARS wins in $n=200,p=100$. In simulation 4 and 5 which has linear and interactive treatment effect, our proposed method dominated. in simulation 6 which has a interactive treatment effect, casual forest dominate in first 2 scenarios and causal BART wins the rest.

Figure 3 is the result of MSE of observational  study simulations 7 to 12. in simulation 7, which has a constant treatment effect, causal BART dominated in all scenarios, and our proposed method performed second to third place. In simulation 8, which has a piecewise interactive treatment effect, causal BART performed best in all scenarios and our proposed method also looks good and performed second place. In simulation 9, which has an additive treatment effect our proposed method dominate in scenario of $n=500$, and causal BART wins in $n=200,p=100$. In simulation 10 and 11 which has linear and interactive treatment effect, our proposed method dominated. In simulation 12 which has a interactive treatment effect, causal BART dominated in all scenarios.

Figure 4 is the result of bias of RCT study simulations 1 to 6. in simulation 1, which has a constant treatment effect, the causal BART dominated in all scenarios, and our proposed method performed second place. In simulation 2, which has a piecewise interactive treatment effect, there is no model that is clearly superior to others. In simulation 3, which has an interactive treatment effect, there is no model that is clearly superior to others. Our proposed method is slightly behind other methods. In simulation 4 which has linear treatment effect, there is no model that is clearly superior to others. In simulation 5 which has a interactive treatment effect, virtual twins is slightly behind other methods. In simulation 6 which has a interactive treatment effect, Our proposed method is slightly behind other methods.

Figure 5 is the result of bias of observational study simulations 7 to 12. in simulation 7, which has a constant treatment effect, causal BART dominated in all scenarios, and our proposed method performed second. In simulation 8, which has a piecewise interactive treatment effect, causal BART performed best in all scenarios. In simulation 9, 10 and 11  which has both linear and interactive treatment effect, our proposed method dominated. In simulation 12, which has an interactive treatment effect, our proposed method and casual BART performed better than the other models. 

From the simulation result, our proposed method is competitive in all cases in this simulation study. In RCT simulations, our proposed method performance is superior to others in simulation 3, 4 and 5 both in MSE and bias. which shows that our proposed method has great fitting for liner models and interactive models. In simulation 1, our proposed model also performed well in MSE and bias, but the causal BART model showed stronger dominance in such a relatively simple model structure. In simulation 2, our proposed model has high bias, because the piecewise function is misspecified for MARS based method. The situation is similar in simulation 6, our proposed model do not workout well in MSE and bias. It might because there is stepwise structure in the treatment effect model, which is misspecified for MARS based methods. And it leads to both high predict error. In observational study simulations, our proposed method performance is similar to RCT simulations. It is more clearly superior to others in simulation 3, 4 and 5. In simulation 6, our proposed model do not workout well in MSE, but has the lowest bias. The model misspecifying still leads to both high variance. But even in this case, our model still has some competitiveness. In the remaining observational simulations, our proposed model shows stronger competitiveness compared to RCT simulations. The result of observational study simulations shows that our propensity adjustment strategy works well as expected. In general, our model can fit various constructed treatment effect models well, and perform extremely well in both additive models and complex nonlinear and linear models. However, when fitting functions with stepwise structure, the predict error will be relatively large, which is caused by the characteristics of the MARS model.

\section{Real Data Example}\label{sec5}
In this section, we adopt a real data from AIDS Clinical Trial Group 175 (ACTG 175) study (Hammer et al., 1996)\cite{Hammer1996} to our proposed methods. In this randomized clinical trial, four arms were included: zidovudine monotherapy, zidovudine plus didanosine, zidovudine plus zalcitabine, and didanosine monotherapy. The objective was to study the efficacy of these treatments in adults infected with human immunodeficiency virus type 1 (HIV-1) whose CD4 cell counts ranged from 200 to 500 per cubic milliliter.  The CD4 T cell count is a essential indicator used to quantify the risk of HIV-1 infection, where a low count implies a higher risk of infection. The R package BART provides the ACTG 175 dataset, which includes 1762 subjects with baseline CD4 cell counts ranging from 200 to 500 per cubic milliliter for analysis. In this papper, We focus on contrast the treatment effect of zidovudine monotherapy versus the combination therapy with zidovudine and didanosine. The control group consisted of 419 subjects receiving zidovudine monotherapy, while the treatment group consisted of 436 subjects receiving combination therapy with zidovudine and didanosine. The outcome was set to the increase or decrease of CD4 T cell count at $20\pm5$ weeks from baseline and represented as ${(y}_{1i}-y_{0i})/y_{0i}$. where $y_{1i}$  represents the CD4 T cell count for patient $ i $ at 20 weeks and $ y_{0i} $ is the baseline CD4 T cell count. The fallowing twelve variables are included: age: age in years at baseline; wtkg: weight in kg at baseline; hemo: hemophilia (0=no, 1=yes); homo: homosexual activity (0=no, 1=yes); drugs: history of intravenous drug use (0=no, 1=yes); z30: zidovudine use in the 30 days prior to treatment initiation (0=no, 1=yes); preanti: number of days of previously received antiretroviral therapy; race: race (0=white, 1=non-white); gender: gender (0=female, 1=male); symptom: symptomatic indicator (0=asymptomatic, 1=symptomatic); cd40: CD4 T cell count at baseline; cd80: CD8 T cell count at baseline. 

By applying the proposed method with selected parameters to the ACTG 175 dataset, we obtained an HTE model of polynomial with 23 terms, as shown in \textbf{Eq.5}. Our proposed method constructs the model in the form of a polynomial, allowing for a direct interpretation of the relationship between the basis function and HTE. Meanwhile, the other widely used treatment effect models like tree methods are constructed as black box estimators.
\textcolor{red}{
\begin{align}
\widehat{HTE}=&-7.583507e^{-2}+5.346738e^{-4}*drugs*[(348-cd40.)]_+-1.422048e^{-6}[cd80-620]_+\nonumber\\\nonumber&-3.899131e^{-7}[cd80-617]_++2.301013e^{-5}[867-cd80.]_+\\\nonumber&+8.172967e^{-4}[(age-37)]_+*[(200-cd40)]_+-7.732336e^{-5}[cd40-470]_+\\\nonumber&-7.126319e^{-6}[66.6792-wtkg]_+*[269-cd40]_+-6.076188e^{-7}[cd80-575]_+ \\\nonumber&+1.193716e^{-3}[643-cd40]_+-5.639245e^{-3}[53.9784-wtkg]_+-3.362105e^{-7}[cd80-626]_+\\\nonumber&-2.398545e^{-4}[wtkg-93.6]_+-4.601273e^{-6}[cd80-605]_+-1.691710e^{-7}[cd80-624]_+\\\nonumber&+7.132453e^{-5}[1718-cd80]_++4.146636e^{-4}[age-32]*[198-cd40]_+\\\nonumber&-8.084559e^{-5}[wtkg-93.4416]_++8.066342e^{-6}[858-cd80]_+\\\nonumber&-1.742378e^{-7}[cd80-622]_++8.290441e^{-6}[868-cd80]_+\\&+2.175676e^{-4}[wtkg-70.9]_+*[240-cd40]_+-1.506107e^{-5}[cd40-373]_+
\end{align}
Here we randomly draft some samples from the data to show how our proposed method interpret the HTE in $\mathbf{Table}$ \ref{tb2}  .
\begin{table}[!ht]
\centering
\caption{ Examples of subjects covariates and estimated HTE}
\resizebox{0.9\textwidth}{!}{
\label{tb2}
\begin{tabular}{l|llllllllllll|l}
\rowcolor[HTML]{C0C0C0} 
id &age &wtkg &hemo &homo &drugs &z30 &race &gender &str2 &symptom &cd40 &cd80 & estimated HTE\\ \hline
11983 & 42 & 76.50 & 0 & 1 & 0 & 0 & 0 & 1 & 0 & 0 & 453 & 600 & 0.2399384\\
90693 & 28 & 73.71 & 0 & 1 & 0 & 0 & 0 & 1 & 0 & 0 & 373 & 631 & 0.3330451\\
110681 & 37 & 65.50 & 0 & 1 & 0 & 1 & 0 & 1 & 1 & 1 & 370 & 1260 & 0.2776932\\
\end{tabular}
}
\end{table}
For subject 11983, we have 
\begin{align*}
\widehat{HTE}_{11983} &= -7.583507e^{-2}+2.301013e^{-5}[867-cd80.]_++1.193716e^{-3}[643-cd40]_+\nonumber\\ &  +7.132453e^{-5}[1718-cd80]_++8.066342e^{-6}[858-cd80]_+\nonumber\\ &  \nonumber+8.290441e^{-6}[868-cd80]_+-1.506107e^{-5}[cd40-373]_+\nonumber\\
&=7.5793e^{-3} - 2.4e^{-6}cd80 + 1.194e^{-3}cd40
\end{align*}
Similarly, we can generate the estimated HTE model for any subjects. According to the definition of hinge function, it is easy to find out that the model of subject is not of a certain subject, but for a subgroup. inside a subgroup with covariant within a specific range, subject will share a same HTE model. which is similar to regression tree, but instead of using a value to represent the HTE of whole subgroup, our proposed method use a model to estimate HTE. From this perspective, our model, while possessing nearly the same level of interpretability as regression trees, offers a greater degree of "personalization."}
\begin{figure}[!ht]
\begin{center}
\includegraphics[width=0.8\linewidth]{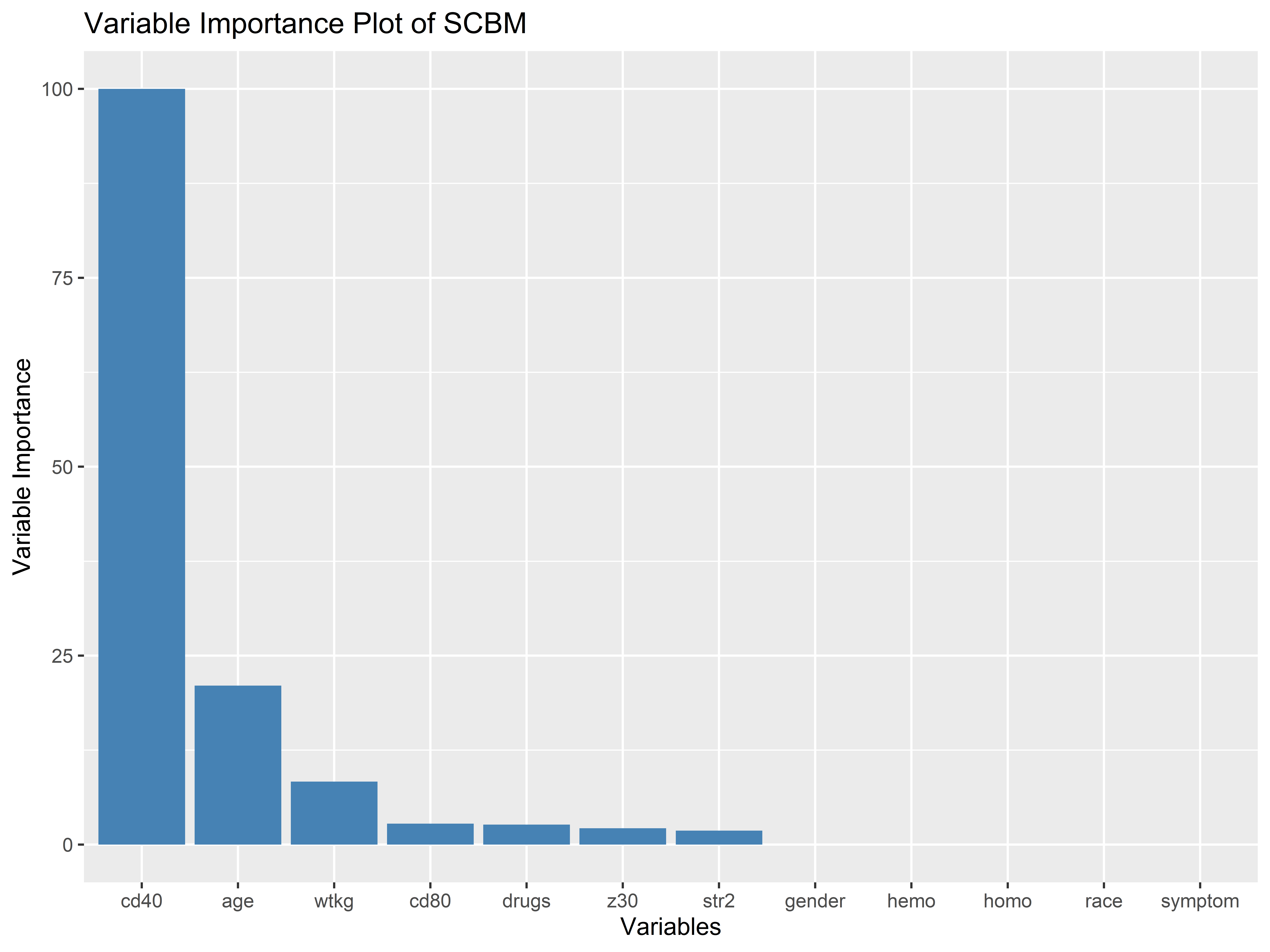}
\caption{Variable importance plots for ACTG175}
\label{fig:3}
\end{center}
\end{figure}

\begin{figure}[ht]
\begin{center}
\includegraphics[width=0.8\linewidth]{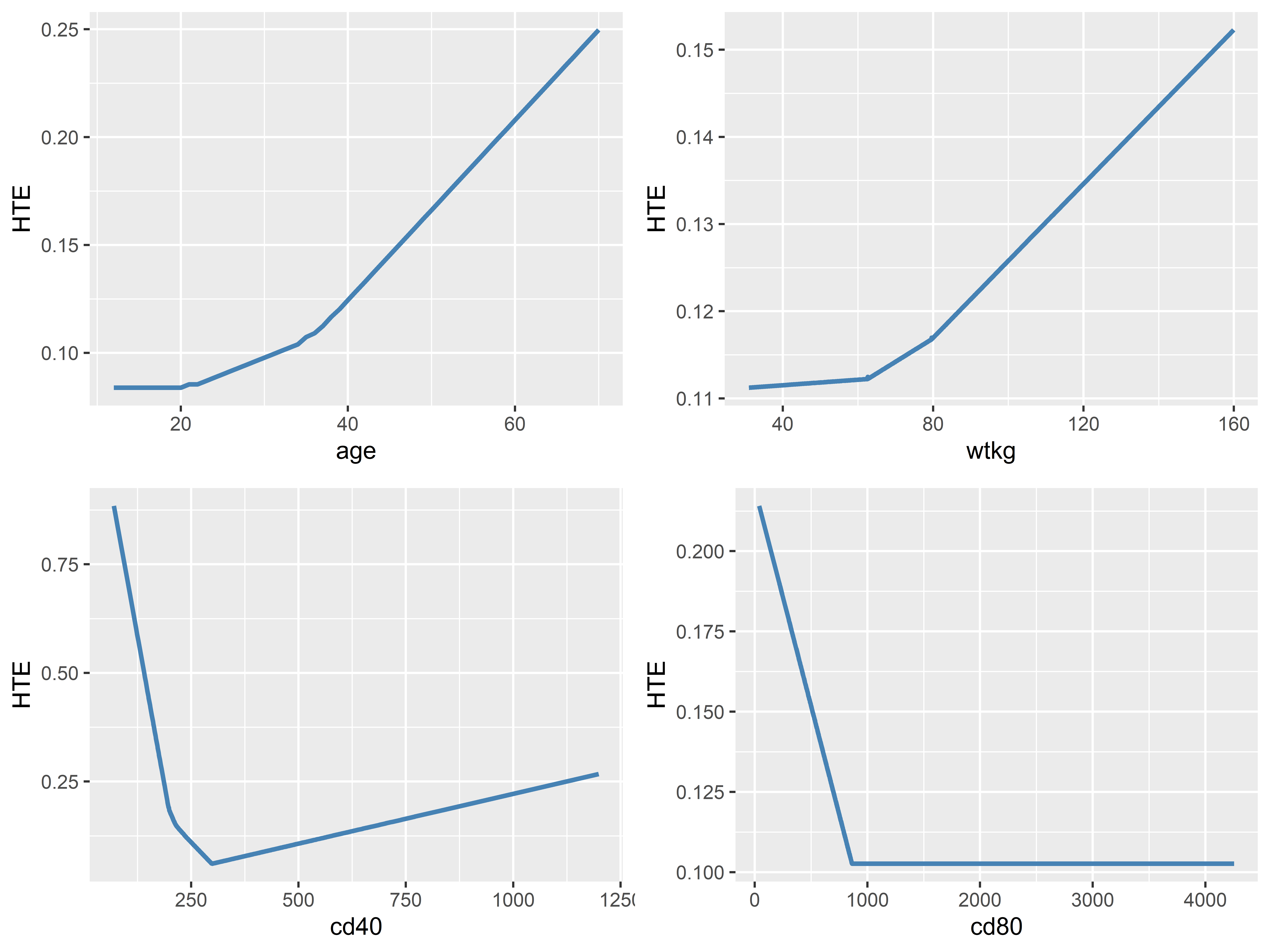}
\caption{Partial dependence plots for ACTG175}
\label{fig:4}
\end{center}
\end{figure}
Figure~\ref{fig:3}  is the variable importance plots (Breiman, 2001)\cite{Breiman2001} for SCBM method. These variable importance plots provided the variable importance of the most important variables in descending order, and the variable with higher variable importance contribute more to the model than the lower ones. Consider $n\times p$ matrix of variables $\bm{x}$ of $n$ samples and a vector of response $\bm{y}$. Let $j=1,\cdots,p$; $i=1,\cdots,n$.  $t_i$ denotes the binary treatment indicator, where $t_i=1$:treatment group; $t_i=0$:control group. Let $S\left(\bm{x}_i,t_i\right)$ denote the proposed method, $ Loss\left(\bm{y},\bm{x},\bm{t}\right)$ denote the sum of square error of the model
\begin{align}  
\widehat{Loss}\left(\bm{y},\bm{x},\bm{t}\right)=\sum_{i=1}^{n}{(y_i-S{\left(\bm{x}_i,t_i\right))}^2}. 
\end{align}

For Initialization, compute the sum of square error of original data $\widehat{Loss}\left(\bm{y},\bm{x},\bm{t}\right)$ with \textbf{Eq.6}. For every element in predictive variable, create matrix $\bm{x}^{-j}$ by removing the $j-th$ variable from the original data. Then the variable-importance $Vimp^j$ is given by
\begin{align}  
\widehat{Vimp_j}=\widehat{Loss}\left(\bm{y},\bm{x}_{-j},\bm{t}\right)-\widehat{Loss}\left(\bm{y},\bm{x},\bm{t}\right).\nonumber
\end{align}
After traverse all the predictor variables, we would have a collection of variable-importance $Vimp=\{ \widehat{Vimp^j}\mid j=1,\cdots,p\}$. Do normalization:
\begin{align}  
\widehat{nomalVimp_j}=\frac{\widehat{Vimp_j}}{\max{\left(Vimp\right)}}\cdot100\%. \nonumber
\end{align}
After all, make the variable-importance plot with $\widehat{nomalVimp_j}$.

Figure~\ref{fig:4} is the partial dependence plot (Friedman, 2001)\cite{Friedman2001} of SCBM, which shows the marginal effect of the most important variables on prediction of HTE. After identifying the variables importance, the next step is to reveal how the expected value of model prediction behaves as a function of the joint values of the input variables. However, this method is limited to low-dimensional parameters. For more than two or three variables of interest, it is more difficult to observe functions of the corresponding high-dimensional parameters. Sometimes, a useful alternative is to view a set of plots, each showing the partial dependence of the approximation of the model on a selected small subset of the input variables.  Similarly, let $S\left(\bm{x}_i,t_i\right)$ denote the model. Create matrix $\bm{x}^{j=c}$ by replacing the $j-th$ variable with elements $c$,  which is an element of the $j-th$ variable. The value of a partial dependence for $S\left(\bm{x}^{j=c}_i,t_i\right)$ is defined as $H_{PD}^j\left(c\right)=E\left(S(\bm{x}^{j=c}_i,t_i)\right).$ 
Partial dependence for $j-th$ variable with elements $c$ can be estimated by 
\begin{align}  
{\hat{H}}_{PD}^j\left(c\right)=\frac{1}{n}\sum_{i=1}^{n}S\left(\bm{x}_i^{j=c},t_i\right).\nonumber
\end{align}

According to the variable importance plot of SCBM, the CD4 T cell counts obviously contribute more to the model than the other variables.The result of variable importance is partially related to the result of Tsiatis and Spanbauer et al.(Spanbauer and Sparapani, 2021; Tsiatis et al., 2008)\cite{Spanbauer2021}\cite{Tsiatis2008}. The CD4 T cells counts are much higher than other variables which indicate that the CD4 T cells count at baseline is the key variables, then is age at baseline, weight in kg at baseline and CD8 T cells count at baseline. From the partial dependence plot, combination regimen evoked greater response on CD4 T cells count at baseline, especially for patient whose CD4 T cells count at baseline below 300.  In addition to that, the partial dependence plots are all positive, which indicating that combination therapy is likely to be more effective than monotherapy. This is also related to the results of Hammer et al. (1996).

\begin{figure}[ht]
\centering
\begin{center}
\begin{minipage}{0.9\textwidth}
\centering
\includegraphics[width=0.8\textwidth]{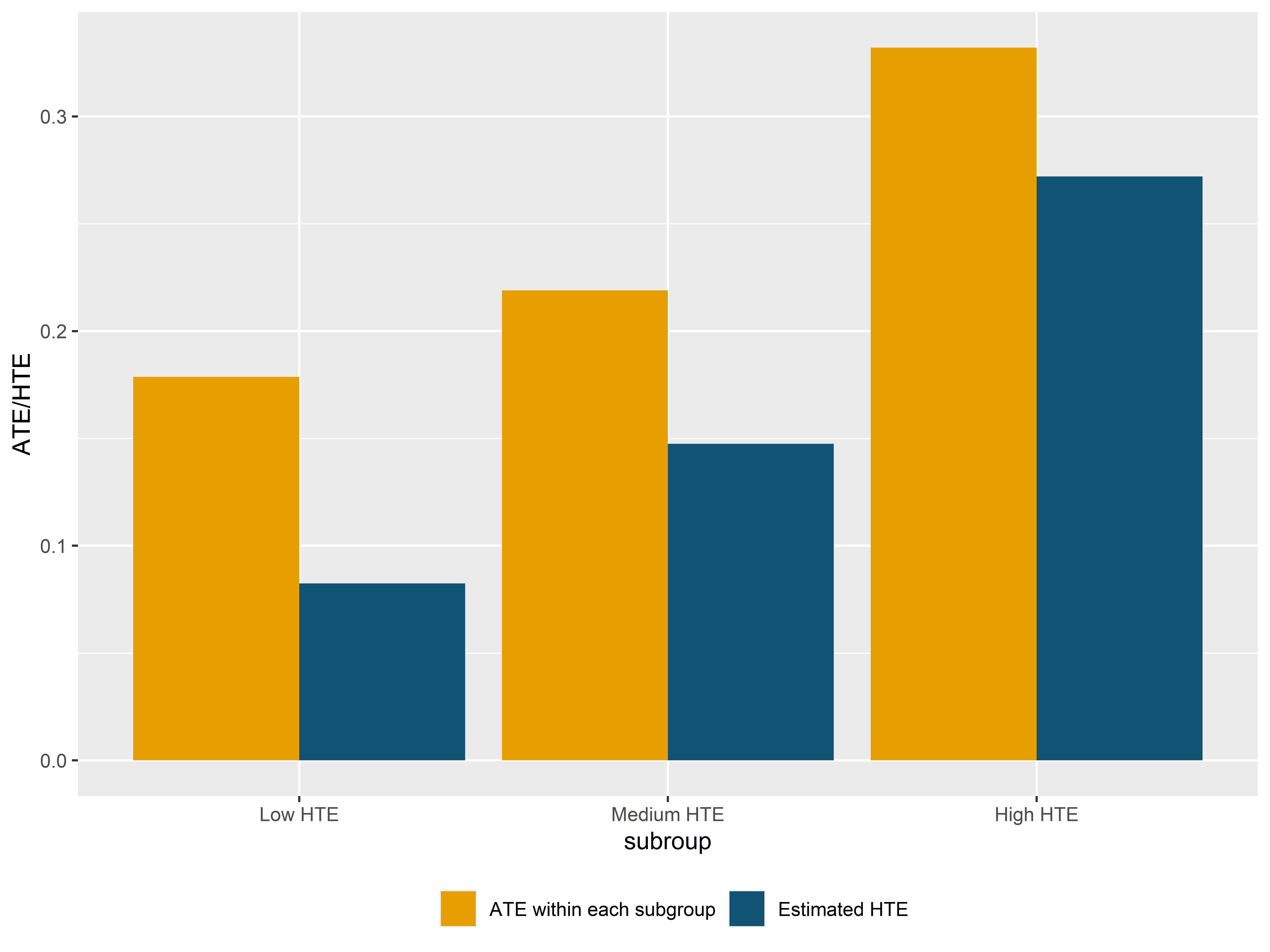}
\captionsetup{format=hang}
\caption{The comparison of estimated HTE of proposed method with ATE within each subgroup at subgroup levels}
\label{fig:5}
\end{minipage}
\end{center}
\end{figure}
%
\begin{figure}[ht]
\centering
\begin{center}
\begin{minipage}{0.9\textwidth}
\centering
\includegraphics[width=0.8\textwidth]{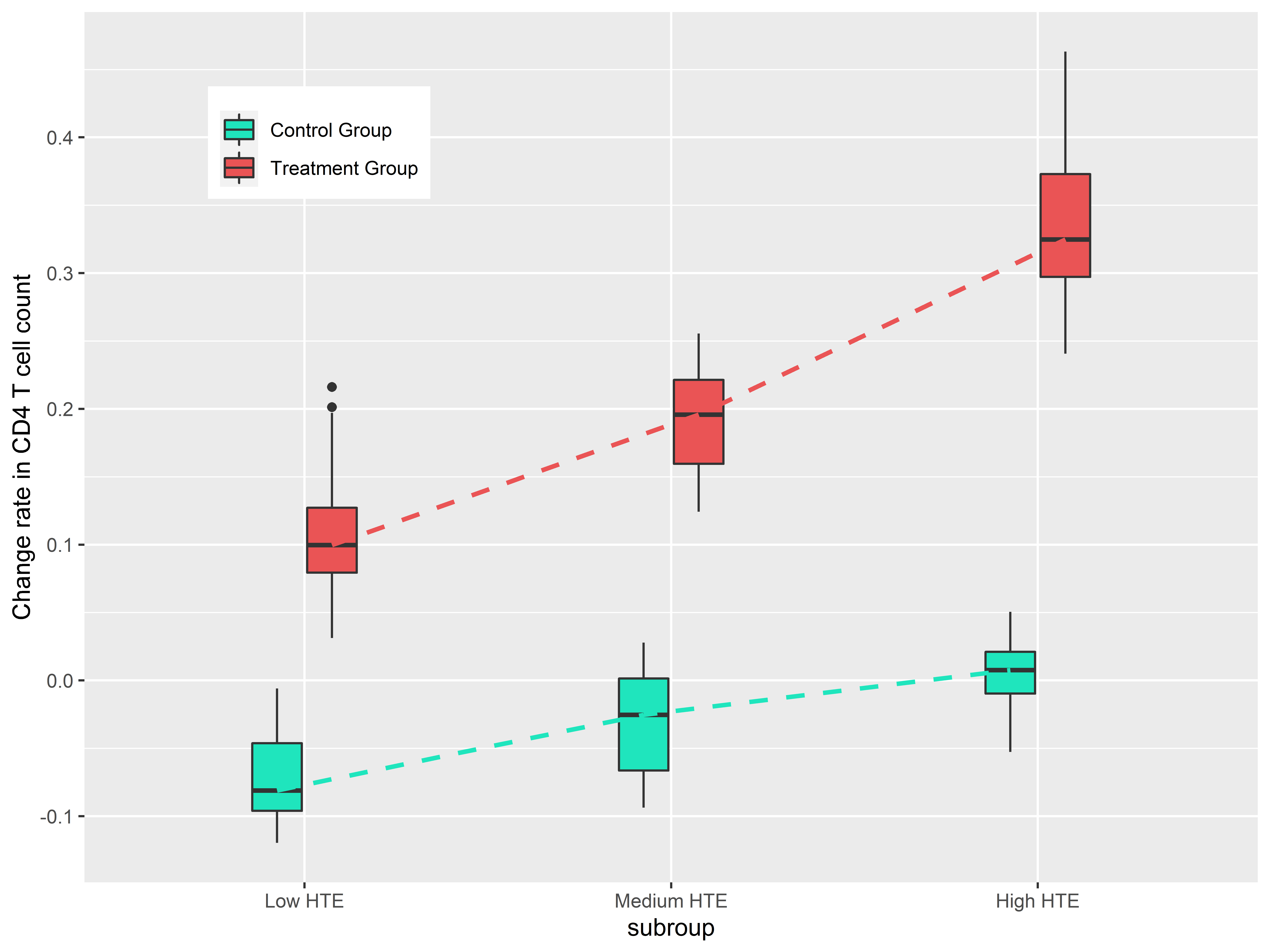}
\captionsetup{format=hang}
\caption{The true outcome within each subgroups}
\label{fig:6}
\end{minipage}
\end{center}
\end{figure}
In real data, unlike in simulation studies, the true HTE for each subject is unknown.To validate the HTE predictive capability of SCBM in real data, we conducted a sampling process. In this process, we randomly selected half of the data and used it as the training set. This sampling process was repeated 100 times. Within each sample, we divided the subjects into nine subgroups of equal size. These subgroups were created based on the ascending order of their corresponding estimated HTE. For each subgroup, we calculated the average treatment effect (ATE) and assumed it to be the true HTE. Additionally, we calculated the average estimated HTE for each subgroup. Finally, we calculated the average estimated HTE and ATE across all the bootstrap samples. This methodology allowed us to assess the validity of SCBM in predicting HTE in real data.
We compared the true HTE and estimated HTE at subgroup level, as shown in Figure~\ref{fig:5}, The proposed method showed a similar trend between the estimated and true HTE at the subgroup level.  The observed outcome within each subgroups is shown in Figure~\ref{fig:6}. We can see that the subgroup with higher HTE demonstrates a greater disparity in patient outcomes between those assigned to the treatment group and those assigned to the control group.
In other words, a high estimated HTE is correlated with a high true HTE and a low estimated HTE is correlated with a low true HTE in the proposed method. By implementing the proposed method, we can identify which individuals benefit from combination therapy and which ones benefit from monotherapy based on their covariate vectors.

\section{Discussion}\label{sec6}
We proposed a MARS based causal method shrinkage causal bagging MARS (SCBM) as an alternative solution to estimate heterogeneous treatment effect problem with MARS method. SCBM is constructed as a shared basis conditional mean regression method where the basis functions are estimated using the transformed outcome bootstrap MARS model and the regression parameters are estimated by group lasso. The transformed outcome MARS generates basis functions that capture the HTE adjusted by propensity score. Then, regularization regression is performed using group lasso regression to regularize the existing basis functions while keeping the equivalence between the experimental and control group basis functions. The proposed method differs from previous tree-based ensemble methods like causal forest and BART in that it creates a polynomial model instead of aggregating multiple tree models. Polynomial model provides SCBM the proposed method a great interpretability.

To validate the performance of the proposed model SCBM, we designed a simulation study to evaluate predictive accuracy. SCBM works very well in randomized control trails study simulations and shows advantage in observational study simulations comparing with causal forest, PTO forest,bagging causal MARS, Virtual twins. We applied SCBM to real dataset ACTG175 to validate the availability of the proposed model SCBM with graphic interpretation, and our proposed method gets the result which is supported by several previous research. And by plotting the comparison of the estimated HTE of the proposed method with the ATE within each subgroup, we can generally assume that a high estimated HTE is correlated with a high true HTE, while a low estimated HTE is correlated with a low true HTE in the proposed method.

In general, the proposed method not only demonstrates outstanding performance in prediction accuracy but also exhibits great interpretability. It is highly suitable for predicting HTE in observational studies. However, there are also some limitations. For instance, when the true model of treatment effects is relatively simple, its capability is not as good as causal BART. Additionally, compared to traditional tree models, it requires more computational resources.
\bibliographystyle{unsrt}  
\bibliography{references}

\end{document}